\documentclass[reprint,aip,jcp,amsmath]{revtex4-1}
\usepackage{graphicx}
\usepackage{txfonts}
\usepackage{color}
\definecolor{myblue}{rgb}{0,0,1}
\usepackage[breaklinks=true,colorlinks=true,linkcolor=myblue,urlcolor=myblue,citecolor=myblue]{hyperref}

\begin{document}

\title{A mean-field treatment of vacuum fluctuations in strong light--matter coupling}
\author{Ming-Hsiu Hsieh}
\altaffiliation{Both authors contributed equally to this work.}
\author{Alex Krotz}
\altaffiliation{Both authors contributed equally to this work.}
\author{Roel Tempelaar}
\email{roel.tempelaar@northwestern.edu}
\affiliation{Department of Chemistry, Northwestern University, 2145 Sheridan Road, Evanston, Illinois 60208, USA}

\begin{abstract}
Mean-field mixed quantum--classical dynamics could provide a much-needed means to inexpensively model quantum electrodynamical phenomena, by describing the optical field and its vacuum fluctuations classically. However, this approach is known to suffer from an unphysical transfer of energy out of the vacuum fluctuations when the light--matter coupling becomes strong. We highlight this issue for the case of an atom in an optical cavity, and resolve it by introducing an additional set of classical coordinates to specifically represent vacuum fluctuations whose light--matter interaction is scaled by the instantaneous ground-state population of the atom. This not only rigorously prevents the aforementioned unphysical energy transfer, but is also shown to yield a radically improved accuracy in terms of the atomic population and the optical field dynamics, generating results in excellent agreement with full quantum calculations. As such, the resulting method emerges as an attractive solution for the affordable modeling of strong light--matter coupling phenomena involving macroscopic numbers of optical modes.
\end{abstract}

\maketitle


Recent years have seen an increasing interest in strong light--matter coupling, especially arising when a quantum emitter in the form of an atom, molecule, or material is embedded in an optical cavity. \cite{ebbesenHybridLightMatter2016} The hybrid light--matter (polariton) states emerging in such implementations have been experimentally demonstrated to allow modulation of the physical and chemical properties of the quantum emitter through cavity tuning, \cite{hertzogStrongLightMatter2019, dunkelbergerVibrationCavityPolaritonChemistry2022} impacting processes such as energy transfer \cite{colesPolaritonmediatedEnergyTransfer2014} and chemical reactivity. \cite{shiEnhancedWaterSplitting2018, thomasTiltingGroundstateReactivity2019} Nevertheless, various open questions evade direct experimental investigations, such as the extent to which such modulation is impacted by collective effects, including collective strong coupling \cite{pinoQuantumTheoryCollective2015} and dark state manifolds \cite{gonzalez-ballesteroUncoupledDarkStates2016}, motivating a surge in theoretical investigations \cite{ruggenthalerQuantumelectrodynamicalLightMatter2018, flickStrongLightmatterCoupling2018, ribeiroPolaritonChemistryControlling2018, liMolecularPolaritonicsChemical2022, sanchez-barquillaTheoreticalPerspectiveMolecular2022, fregoniTheoreticalChallengesPolaritonic2022, lindoyQuantumDynamicsVibrational2022}.

Due to the considerable computational cost associated with modelling the optical field fully quantum mechanically, many theoretical investigations have adopted the Jaynes--Cummings \cite{jaynesComparisonQuantumSemiclassical1963} and Tavis--Cummings models \cite{tavisExactSolutionMolecule1968}, wherein the field is represented by only a few or even a single mode. In addition to a neglect of the macroscopic number of optical modes and its potential to influence polariton dynamics, \cite{hoffmannEffectManyModes2020} such models commonly ignore the finite optical wavelength and the speed of light. All of these factors can conveniently be represented within a classical field representation. Indeed, such is done in the popular finite difference time-domain (FDTD) \cite{yeeNumericalSolutionInitial1966} and the finite element method (FEM) \cite{hrennikoffSolutionProblemsElasticity2021, courantVariationalMethodsSolution1943}, but without invoking a quantum-mechanical description of quantum emitters.

A straightforward way to unify classical field representations with a quantum-mechanical description of quantum emitters is offered by mixed quantum--classical dynamics. This class of techniques has received attention as a means to model strong light--matter coupling, with efforts particularly addressing its ability to describe the spontaneous emission of a quantum emitter prepared in an excited state, and embedded in a cavity at zero temperature. \cite{hoffmannCapturingVacuumFluctuations2019, hoffmannBenchmarkingSemiclassicalPerturbative2019, liQuasiclassicalModelingCavity2020, SallerBenchmarkingQuasiclassicalMapping2021} Accounting for spontaneous emission within a fully classical theory is straightforward in principle, \cite{millerClassicalSemiclassicalTheory1978} but it poses an interesting challenge once the classical approximation is exclusively retained for the optical field, since in that case the interaction with an excited-state quantum emitter vanishes at low temperatures due to the absence of vacuum fluctuations. \cite{LiMixedQuantumClassical2018} Although this shortcoming can be remedied through incorporation of additional relaxation pathways, \cite{chenEhrenfestDynamicsMixed2019} other efforts have instead resorted to related semiclassical approaches where quantum effects can be directly incorporated in the classical degrees of freedom. \cite{liQuasiclassicalModelingCavity2020, SallerBenchmarkingQuasiclassicalMapping2021} 

\begin{figure}
  \includegraphics{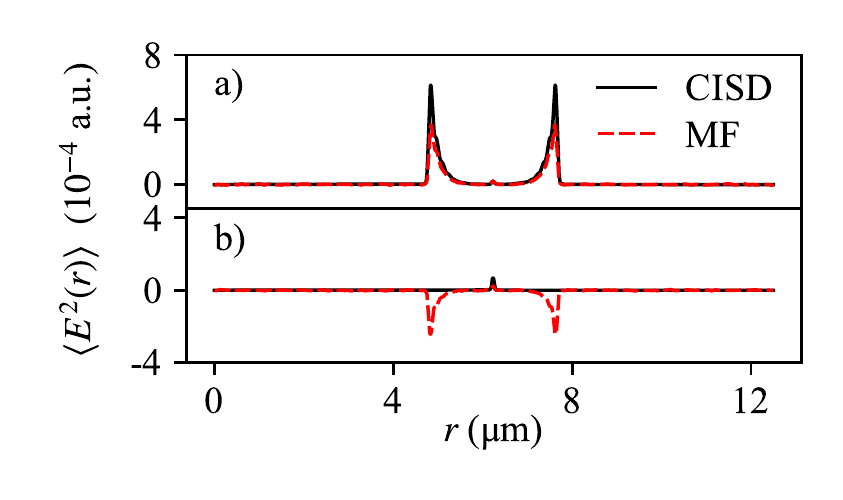}
  \caption{Optical field intensity shortly after spontaneous emission by an atom located at the center of an optical cavity, initiated in a) its excited state and b) its ground state. Shown are results at  $t=200$~a.u., calculated with CISD (black solid curve) and with MF dynamics (red dashed curve). Each trace was smoothed by a $0.05$ $\mathrm{\mu}$m moving average for ease of demonstration.}
  \label{fig_1}
\end{figure}

Recently, Hoffmann \emph{et al.}~reported promising results for the spontaneous emission of a cavity-embedded atom by applying mean-field (MF) mixed quantum--classical dynamics while capturing the optical vacuum fluctuations by sampling classical modes from a Wigner distribution, which retains finite energy at zero temperature. \cite{hoffmannCapturingVacuumFluctuations2019, hoffmannBenchmarkingSemiclassicalPerturbative2019} The accuracy reached in such an implementation (henceforth simply referred to as MF dynamics) was shown to be qualitatively similar compared to more elaborate semiclassical methods, while being superior compared to trajectory-based surface hopping methods \cite{hoffmannBenchmarkingSemiclassicalPerturbative2019}. Combined with its unrivaled simplicity and affordability, this suggests that MF dynamics may be particularly suitable for describing strong light--matter coupling phenomena.

The success of MF dynamics in describing strong light--matter coupling is underscored by Fig.~\ref{fig_1}(a) which reproduces a result previously reported by Hoffmann \emph{et al.}  \cite{hoffmannCapturingVacuumFluctuations2019} (the applied model and parameters are the same as employed in the present study, and will be elaborated upon below). Here, a comparison is drawn between MF dynamics and the near-exact configuration interaction singles and doubles (CISD) for the optical field intensity following spontaneous emission from an atom positioned at the center of the cavity in the zero temperature limit. Indeed, the qualitative agreement between the two methods is seen to be extremely promising, although a comparatively reduced amplitude of the emitted field for MF dynamics reflects incomplete emission and a concomitant incomplete relaxation of the quantum emitter into its ground state. Perhaps a more pertinent shortcoming of MF dynamics, however, is exemplified in Fig.~\ref{fig_1}(b), presenting newly-obtained results for the same cavity-embedded atom but initiated in its ground state. Here, the optical field is seen to feature pronounced negative wavefronts that propagate away from the atom, which is indicative of an unphysical energy transfer from the optical vacuum field to the ground state atom. This unphysical energy transfer renders MF dynamics inapplicable to strong light--matter phenomena which typically involve a multitude of ground state quantum emitters that would all contribute negative wavefronts to the optical field.

It is worth emphasizing that this unphysical energy transfer is not related to the lack of detailed balance that MF dynamics is known to suffer from \cite{parandekarDetailedBalanceEhrenfest2006,parandekarMixedQuantumclassicalEquilibrium2005}, since it occurs much faster than the timescale at which the atom--cavity system comes to equilibrium. Instead, it is due to the light--matter interaction being allowed to remove energy from the optical vacuum fluctuations once the quantum-mechanical Fock basis of the optical modes is replaced by finite-energy classical oscillators. This, however, implies that by appropriately modifying the light--matter interaction, such unphysical energy transfer could be remedied while still retaining a simple and affordable MF description.
 
In this Letter, we propose a pragmatic solution to the unphysical energy transfer from optical vacuum fluctuations to the ground state of the quantum emitter within MF dynamics. This solution relies on the inclusion of two sets of classical coordinates to represent vacuum fluctuations and the thermal contribution to the optical field, respectively. For the vacuum fluctuations, we then subtract the portion of the light--matter interaction weighted by the ground state population of the quantum emitter. Interestingly, in addition to resolving the unphysical transfer of energy, this approach turns out to radically improve the overall accuracy of MF dynamics in describing recurrent energy exchange between an atom and the optical field of a cavity, thereby significantly enhancing the applicability of simple and inexpensive MF techniques to transiently model strong light--matter coupling phenomena involving macroscopic numbers of optical modes.


We consider an atom positioned at the center of a one-dimensional optical cavity, and represented by an electronic two-level system governed by the Hamiltonian
    \begin{eqnarray}
    \hat{H}_{\mathrm{{A}}}=\epsilon_{\mathrm{e}}\hat{c}_{\mathrm{e}}^{\dagger}\hat{c}_{\mathrm{e}}+\epsilon_{\mathrm{g}}\hat{c}_{\mathrm{g}}^{\dagger}\hat{c}_{\mathrm{g}}.\label{eq:HA_1}
    \end{eqnarray}
Here, $\hat{c}^{(\dagger)}_{\mathrm{g}}$ and $\hat{c}^{(\dagger)}_{\mathrm{e}}$ represent the annihilation (creation) operators for the atomic ground and excited states, respectively, and $\epsilon_{\mathrm{e}}$ and $\epsilon_{\mathrm{g}}$ are the associated energies. A single relevant polarization direction of the optical field is represented by a collection of harmonic modes, the Hamiltonian of which is given by
\begin{eqnarray}
\hat{H}_{\mathrm{F}}=\frac{1}{2}\sum_\alpha\left(\hat{P}_\alpha^2+\omega_\alpha^2\hat{Q}_\alpha^2\right).\label{eq:HF_1}
\end{eqnarray}
Here, $\alpha$ labels the modes, the frequency of which are given by $\omega_{\alpha}=\pi c_0 \alpha L^{-1}$, where $c_0$ and $L$ are the speed of light and the cavity length, respectively. Furthermore, $\hat{Q}_{\alpha}$ and $\hat{P}_{\alpha}$ represent the position and momentum operators for mode $\alpha$, which physically are related to the electric and magnetic contributions to the optical field, respectively \cite{pellegriniOptimizedEffectivePotential2015}. The atom-field interaction,\footnote{We omit the self-polarization term which for the model at hand amounts to a shift of the quantum emitter energy levels that can straightforwardly be absorbed in $\epsilon_\mathrm{e}$ and $\epsilon_\mathrm{g}$, and which does not meaningfully alter the dynamics. When representing the quantum emitter by three or more levels, inclusion of the self-polarization term is generally required \cite{hoffmannCapturingVacuumFluctuations2019, hoffmannEffectManyModes2020, liMolecularPolaritonicsChemical2022}}
    \begin{eqnarray}
    \hat{H}_{\mathrm{AF}}=\sum_{\alpha}\omega_{\alpha}\lambda_{\alpha}\mu\left(\hat{c}_{\mathrm{e}}^{\dagger}\hat{c}_{\mathrm{g}}+\hat{c}_{\mathrm{g}}^{\dagger}\hat{c}_{\mathrm{e}}\right)\hat{Q}_{\alpha}, \label{eq:H_AF_quant}
    \end{eqnarray}
can be derived from the Pauli-Fierz Hamiltonian within the dipole approximation. \cite{ruggenthalerQuantumelectrodynamicalLightMatter2018,flickCavityBornOppenheimer2017} Here, $\mu$ is the ground-to-excited state transition dipole moment of the atom, assumed to be aligned parallel to the optical polarization direction. With the atom centered at the cavity, its coupling strength to the optical field is given by
\begin{eqnarray}
\lambda_{\alpha}=\sqrt{\frac{2}{\epsilon_0 L}}\sin\left({\frac{\pi\alpha}{2}}\right),
\end{eqnarray}
where $\epsilon_0$ is the vacuum permittivity. For a detailed discussion of this same model we refer the reader to Refs.~\citenum{hoffmannCapturingVacuumFluctuations2019} and \citenum{flickAtomsMoleculesCavities2017}.
    
Conventionally, a mixed quantum--classical approximation to Eq.~\ref{eq:H_AF_quant} amounts to substituting the optical position and momentum operators by their respective classical coordinates, $\hat{Q}_{\alpha}\rightarrow Q_{\alpha}$ and $\hat{P}_{\alpha}\rightarrow P_{\alpha}$. The dynamics of these coordinates is then governed by Hamilton's equations of motion, which correspond to a mode-resolved form of Maxwell's equations. Importantly, however, these equations involve a force term arising from the ``feedback'' of the quantum state of the atom, given by
\begin{eqnarray}
F_{\alpha}=-\langle\psi\vert \nabla_{Q_{\alpha}} \hat{H}_{\mathrm{AF}} \vert\psi\rangle.
\end{eqnarray}
Within MF dynamics, the feedback state takes the form $\vert\psi\rangle=c_{\mathrm{e}}\vert \mathrm{e}\rangle + c_{\mathrm{g}}\vert \mathrm{g}\rangle$, where $c_{\mathrm{e}(\mathrm{g})}$ are complex-valued expansion coefficients for the excited (ground) state which are propagated through the time-dependent Schr\"odinger equation,
    \begin{eqnarray}
    \vert\dot{\psi}\rangle = -\frac{i}{\hbar}\left(\hat{H}_{\mathrm{A}}+\hat{H}_{\mathrm{AF}}\right)\vert\psi\rangle.\label{eq:tdse}
    \end{eqnarray}
The prevailing means to incorporate vacuum fluctuations of the optical field within such a mixed quantum--classical framework is to sample $Q_{\alpha}$ and $P_{\alpha}$ from a Wigner distribution, so that it statistically reproduces the ground-state wavefunction of the optical field at zero temperature while asymptotically approaching the Boltzmann distribution at high temperatures.

In our approach, we explicitly differentiate between the contributions dominating in the zero and high temperature limits by introducing two sets of position and momentum coordinates for each mode. The first set, $\tilde{Q}_{\alpha}$ and $\tilde{P}_{\alpha}$, is taken to represent vacuum fluctuations and is sampled from a Gaussian distribution corresponding to the ground-state wavefunction of the optical field in Wigner phase-space. The second set, $Q_\alpha$ and $P_\alpha$, is taken to represent thermal fluctuations and is sampled from a Boltzmann distribution. This allows light--matter interactions involving vacuum fluctuations to be formally distinguished from those involving thermal fluctuations. When considering the trivial expression $\tilde{Q}_{\alpha} = \rho_\mathrm{e} \tilde{Q}_{\alpha} + \rho_\mathrm{g} \tilde{Q}_{\alpha}$, where $\rho_\mathrm{e(g)} = \vert c_{\mathrm{e(g)}}\vert ^2$ is the population of the atomic excited (ground) state, it can be appreciated that it is the contribution $\rho_\mathrm{g} \tilde{Q}_{\alpha}$ that leads to the unphysical energy transfer from the vacuum fluctuations of the optical field to the atomic ground state. This contribution can be straightforwardly and pragmatically decoupled by subtracting it from the mixed quantum--classical atom--field Hamiltonian, yielding
    \begin{eqnarray}
    \hat{H}^{\mathrm{DC}}_{\mathrm{AF}}=\sum_{\alpha}\omega_{\alpha}\lambda_{\alpha}\mu\left(Q_{\alpha}+(1-\rho_{\mathrm{g}})\tilde{Q}_{\alpha}\right)\left(\hat{c}_{\mathrm{e}}^{\dagger}\hat{c}_{\mathrm{g}}+\hat{c}_{\mathrm{e}}\hat{c}_{\mathrm{g}}^{\dagger}\right)\label{eq:new_Haf}.
    \end{eqnarray}
This interaction forms the basis of our approach, which we refer to as ``decoupled'' mean-field (DC-MF) dynamics.

We note that the above approach effectively introduces a functional dependence of the atom--field Hamiltonian on the quantum state of the quantum emitter, invoking a linear scaling of the vacuum-driven light--mater interaction terms with the excited state population. The applied linear scaling can be regarded as a first-order approximation to the exact form of this functional dependence, a derivation of which is nontrivial within the mixed quantum--classical formalism. It would be of interest to further explore this topic, perhaps through an approximate analysis, although the favorable results reported in the present study (\textit{vide infra}) suggest the applied first-order approximation to hold well.

Apart from the atom--field interaction and initial conditions, the presence of two sets of position and momentum coordinates leaves DC-MF dynamics virtually identical to conventional MF dynamics. Both sets of coordinates contribute to the classical field Hamiltonian, 
    \begin{eqnarray}
    H_\mathrm{F}=\frac{1}{2}\sum_{\alpha}\left(P_{\alpha}^{2}+\omega_{\alpha}^{2}Q_{\alpha}^{2}+\tilde{P}_{\alpha}^{2}+\omega_{\alpha}^{2}\tilde{Q}_{\alpha}^{2}\right),
    \end{eqnarray}
while their time evolution is governed by Hamilton's equations,
    \begin{eqnarray}
    \dot{Q}_{\alpha}=P_{\alpha}\qquad\dot{P}_{\alpha}=-\omega_{\alpha}^{2}Q_{\alpha}+F_{\alpha} \\
    \dot{\tilde{Q}}_{\alpha}=\tilde{P}_{\alpha}\qquad\dot{\tilde{P}}_{\alpha}=-\omega_{\alpha}^{2}\tilde{Q}_{\alpha}+\tilde{F}_{\alpha}\nonumber.
    \end{eqnarray}
Here, $F_{\alpha}$ and $\tilde{F}_{\alpha}$ are the feedback forces given by
    \begin{eqnarray}
    F_{\alpha}=-\left\langle \psi \left\vert\frac{\partial \hat{H}_{\mathrm{AF}}}{\partial Q_{\alpha}}\right\vert\psi\right\rangle \qquad \tilde{F}_{\alpha}=-\left\langle \psi \left\vert\frac{\partial \hat{H}_{\mathrm{AF}}}{\partial \tilde{Q}_{\alpha}}\right\vert\psi\right\rangle,
    \end{eqnarray}
which depend on the state of the quantum subsystem, $\vert\psi\rangle$, with dynamics given by Eq.~\ref{eq:tdse}.

It is worth noting that the explicit time-dependence associated with $\rho_\mathrm{g}$ in Eq.~\ref{eq:new_Haf} breaks the energy conserving nature of Hamilton's equations unless the derivatives of $\rho_\mathrm{g}$ with respect to the classical coordinates are accounted for. A closed-form solution of the terms necessary to rigorously conserve energy may not exist due to the interdependence of the quantum state and the classical coordinates. While it may be possible to approximate the influence of this solution, the negligible deviation of the total energy (approximately 1\% of the energy transferred between the quantum and classical subsystems) renders the current form of Eq.~\ref{eq:new_Haf} sufficient for the present study.

The evaluation of observables within our formulation is performed separately over each set of coordinates while subtracting the initial contribution from the vacuum fluctuations. Accordingly, the expectation value of the optical field intensity is given by
    \begin{eqnarray}
    \langle E^{2} (r) \rangle&=&2\sum_{\alpha,\beta}\bigg[\hbar^{-1}\sqrt{\omega_{\alpha}\omega_{\beta}}\zeta_{\alpha}(r)\zeta_{\beta}(r)
    \left(\langle Q_{\alpha}Q_{\beta}\rangle +\langle\tilde{Q}_{\alpha}\tilde{Q}_{\beta}\rangle\right)\bigg] \nonumber \\
    &&
    - \sum_{\alpha}\zeta_{\alpha}^{2}(r),
    \end{eqnarray}
where
\begin{eqnarray}
\zeta_{\alpha}(r)=\sqrt{\frac{\hbar\omega_{\alpha}}{\epsilon_0L}}\sin\left(\frac{\pi \alpha}{L} r\right)
\end{eqnarray}
is the mode function at the position $r$ within the cavity. The expectation value of the photon number is given by
    \begin{eqnarray}
    \langle N_{\mathrm{ph}} \rangle=\frac{1}{2}\sum_{\alpha}\left(\frac{\langle P_{\alpha}^2 \rangle+ \langle\tilde{P}_{\alpha}^2\rangle}{\hbar\omega_{\alpha}} + \frac{\omega_{\alpha} (\langle Q_{\alpha}^2\rangle+\langle\tilde{Q}_{\alpha}^2\rangle)}{\hbar} -1 \right).
    \end{eqnarray}

All calculations presented in this work employ the same parameters as Hoffmann \emph{et al.} \cite{hoffmannCapturingVacuumFluctuations2019}, which in turn were based on Ref.~\citenum{flickAtomsMoleculesCavities2017}. The atom is represented by the first two energy levels of a one-dimensional hydrogen atom \cite{suModelAtomMultiphoton1991}, yielding $\epsilon_{\mathrm{g}}= -0.6738$ a.u., $\epsilon_{\mathrm{e}}=-0.2798$ a.u., and  $\mu=1.034$ a.u., and the length of the cavity is taken to be $L=2.362\times 10^{5}$ a.u.\ $=12.5$ $\mathrm{\mu}$m. The optical field is represented by its 400 lowest-energy modes, several of which are nearly resonant with the atomic transition since the energy spacing between modes is small compared to the transition energy.\footnote{We take $\alpha$ to range from $1$ to $400$ and neglect modes with even $\alpha$ which are decoupled ($\lambda_{\alpha}=0$) from the quantum emitter at the center of the cavity due to symmetry.} The temperature was taken to be zero for all reported calculations, and convergence with respect to the time integration step was assured in all cases.

For MF and DC-MF dynamics ensemble averages were taken over $10^{6}$ trajectories, each sampled from the relevant zero-temperature classical distribution, meaning that for DC-MF dynamics $Q_\alpha$ and $P_\alpha$ were initiated as zero while $\tilde{Q}_\alpha$ and $\tilde{P}_\alpha$ were drawn from a Gaussian distribution, whereas for MF dynamics $Q_\alpha$ and $P_\alpha$ were drawn from a zero-temperature Wigner distribution (corresponding to the same Gaussian profile as used for DC-MF). In addition, we carried out CISD calculations at zero temperature (truncating the photonic mode bases to double excitations), where we employed the same Hamiltonian (Eqs.~\ref{eq:HA_1}--\ref{eq:H_AF_quant}) and parameters. For CISD calculations initialized in the atomic excited state we applied the initial condition $\vert\psi\rangle=N\hat{c}_{\mathrm{e}}^{\dagger}\vert \phi_{0}\rangle$, where $\vert\phi_{0}\rangle$ is the lowest-energy eigenstate of the combined atom--cavity Hamiltonian within the CISD basis, and where $N$ is a normalization constant. Accordingly, the atomic ground state initial condition was taken to be $\vert\psi\rangle=\vert\phi_{0}\rangle$. For MF and DC-MF dynamics initiating in the ground and excited states we applied the initial conditions $\vert\psi\rangle=\vert\mathrm{g}\rangle$ and $\vert\psi\rangle=\vert \mathrm{e}\rangle$, respectively. 

Within DC-MF dynamics, calculations initiating in the atomic ground state rigorously yield vanishing feedback forces, and thus a complete absence of dynamics, which is therefore not shown here. For atomic excited-state initial conditions, on the other hand, we were surprised to see that DC-MF dynamics not only retains the favorable properties of MF dynamics, but also radically improves its accuracy (\emph{vide infra}).

 \begin{figure}[t]
  \includegraphics{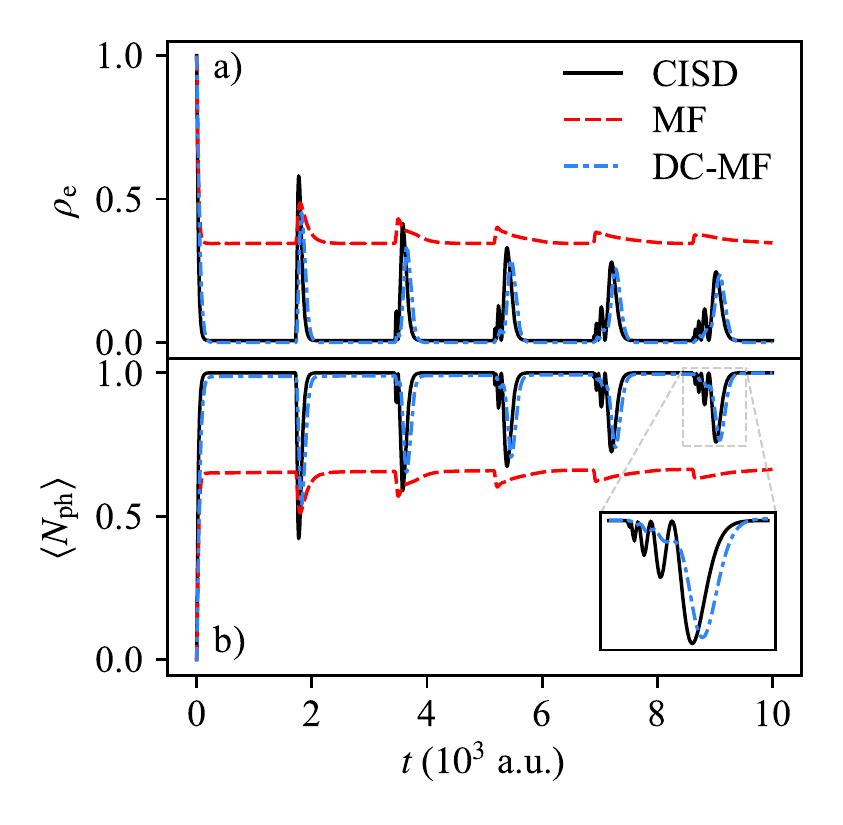}
  \caption{a) Excited-state population of the atom and b) cavity photon number calculated with CISD (black solid curve), MF dynamics (red dashed curve), and DC-MF dynamics (blue dash-dotted curve).}
  \label{fig_2}
  \end{figure}

Fig.~\ref{fig_2}(a) shows the time-dependent population of the atomic excited state, $\rho_{\mathrm{e}}$, resulting from MF dynamics, DC-MF dynamics, and CISD, whereas the corresponding time-dependent cavity photon number, $\langle N_{\mathrm{ph}}\rangle$, is shown in Fig.~\ref{fig_2}(b). The MF results have been previously discussed by Hoffmann \emph{et al.} \cite{hoffmannCapturingVacuumFluctuations2019,hoffmannBenchmarkingSemiclassicalPerturbative2019}, but will be briefly reiterated in the following. Immediately upon initiation, the atom undergoes spontaneous emission into the cavity leading to a relaxation into its ground state, which is manifested as a sharp decay of $\rho_{\mathrm{e}}$ and a concomitant rise in $\langle N_{\mathrm{ph}}\rangle$. Whereas in CISD this relaxation process is seen to be almost complete, with $\rho_{\mathrm{e}}$ decaying to $\sim0$ and $\langle N_{\mathrm{ph}}\rangle$ rising to $\sim1$, a sizable excited-state population is retained within MF dynamics. As time evolves, this population remains reasonably stable, while peaks appearing at time intervals of approximately $1725$ a.u.~correspond to the emitted optical wavefront impinging on the atom, resulting in re-absorption and re-emission. Notably, only a small fraction of the optical field becomes re-absorbed in MF dynamics, leading to peaks that deviate from the CISD results both in amplitude and in shape.

Interestingly, the decoupling of the optical vacuum fluctuations from the atomic ground state results in a near-complete emission from the excited-state atom into the cavity field, as can be seen from the DC-MF results in Fig.~\ref{fig_2}(a), with $\rho_{\mathrm{e}}$ and $\langle N_{\mathrm{ph}}\rangle$ in excellent agreement with the CISD result. Even more promising is that the re-absorption of the emitted optical wavefront by the atom is reproduced almost quantitatively by DC-MF dynamics, with an impressive agreement even after multiple re-absorption and re-emission events. These features are a hallmark of the strong coupling regime, \cite{friskkockumUltrastrongCouplingLight2019} and are a direct result of the neglect of cavity dissipation in the model under study. Comparing the MF and DC-MF methods, the latter suppresses the interference between vacuum contributions and thermal contributions to the light--matter Hamiltonian, yielding a markedly improved reproduction of these features. Particularly notable is the formation of time-dependent fringes in both $\rho_{\mathrm{e}}$ and $\langle N_{\mathrm{ph}}\rangle$ in the course of multiple re-absorption and re-emission events. Notably, however, at short times MF and DC-MF dynamics yield identical results because their respective interaction terms are identical at $t=0$ and remain similar until $\rho_{\mathrm{g}}$ becomes appreciable. 

\begin{figure}[b!]
  \includegraphics{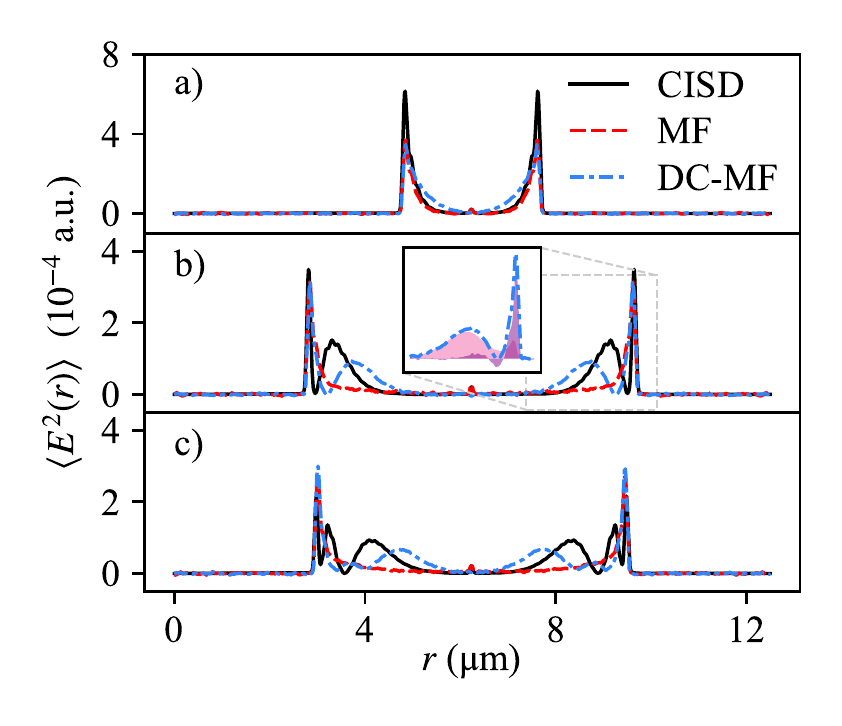}
  \caption{Optical field intensity at  a) $t = 200$~a.u., b) $t = 2200$~a.u., and c) $t = 3900$~a.u., calculated with CISD (black solid curve), MF dynamics (red dashed curve), and DC-MF dynamics (blue dash-dotted curve). The inset shows the contribution of the $Q_{\alpha}$ (pink shade) and $\tilde{Q}_{\alpha}$ (purple shade) coordinates to the field intensity. Each trace was smoothed by a $0.05$ $\mathrm{\mu}$m moving average for ease of demonstration.}
\label{fig_3}
\end{figure}  

Shown in Fig.~\ref{fig_3} is the optical field intensity at different time intervals after initiating the atom in its excited state, resulting from MF dynamics, DC-MF dynamics, and CISD. Interestingly, as seen in Fig.~\ref{fig_3}(a), MF dynamics best reproduces the CISD result at a time $t=200$ a.u., which is immediately after the first emission event. However, it fails to reproduce a lagging portion of the field present in the CISD results after the first and second reabsorptions, as seen in Figs.~\ref{fig_3}(b) and (c). These lagging wavefronts are qualitatively reproduced within DC-MF dynamics, yielding an overall field profile in better agreement with CISD. Shown as an inset in Fig.~\ref{fig_3}(b) is the optical field intensity of DC-MF dynamics differentiated into contributions from $Q_\alpha$ and $\tilde{Q}_\alpha$, where it can be seen that the former induces the shoulders of the wave packet while the latter induces the leading peaks.

One aspect of the CISD results that is absent in DC-MF dynamics is the ground-state contribution to the hybridization of the cavity field with the atomic states induced by counter-rotating terms, which has previously been discussed in the literature. \cite{hoffmannCapturingVacuumFluctuations2019, hoffmannBenchmarkingSemiclassicalPerturbative2019,liQuasiclassicalModelingCavity2020} This hybridization locally enhances the field intensity, which is manifested as a stationary and positive peak at the position of the atom (center of the cavity) in Figs.~\ref{fig_1} and \ref{fig_3}, with a signal amplitude approximately one order of magnitude smaller than that of the emitted wavefronts. The absence of this feature and the underlying hybridization within DC-MF dynamics is reflective of the rigorous decoupling of the atomic ground state from the optical field in the absence of thermal fluctuations. Notably, this feature is present in MF dynamics, where the atomic ground state is allowed to couple to the optical field at all times. However, Fig.~\ref{fig_1}(b) shows that its intensity is underestimated in case when the atom is initialized in the ground state, while previous studies have shown MF dynamics to overestimate its intensity in the case of two-photon emission processes \cite{hoffmannBenchmarkingSemiclassicalPerturbative2019}, suggesting that the hybridization between the atomic ground state and the cavity field is only qualitatively accounted for within MF dynamics.


In summary, we have introduced modifications to MF mixed quantum--classical dynamics enabling the accurate modeling of spontaneous emission by an atom embedded in an optical cavity. Within our method, vacuum fluctuations of the optical field are formally separated from thermal contributions, allowing us to specifically and straightforwardly decouple the light--matter interaction between vacuum fluctuations and the atomic ground state. The resulting DC-MF dynamics is shown to rigorously prevent energy transfer out of vacuum fluctuations, that otherwise manifest as negative intensities in the optical field of the cavity. Moreover, it has the added benefit of radically improving the accuracy of MF modeling for spontaneous emission of a cavity-embedded excited state atom in terms of reproducing the transient atomic populations and the optical field, yielding results in excellent agreement with CISD calculations.

It is worth noting that the agreement between DC-MF results and CISD calculations is even better when the latter is taken within the rotating wave approximation (not shown) \cite{hoffmannBenchmarkingSemiclassicalPerturbative2019}, which seems to suggest that this approximation is related to the decoupling procedure utilized in our method. It should be pointed out, however, that the mixed quantum--classical Hamiltonian used to obtain the DC-MF interaction includes both rotating and counter-rotating terms, and thus is not formally within the rotating wave approximation. We therefore expect DC-MF dynamics to be applicable to the ultra-strong coupling regime (where the rotating wave approximation breaks down) while neglecting some of the ground-state phenomena \cite{friskkockumUltrastrongCouplingLight2019}, which makes for an interesting topic of future inquiry. Furthermore, a multi-atom generalization can be straightforwardly achieved by separately decoupling each atomic ground state. Further future efforts will be geared towards benchmarking DC-MF dynamics for more elaborate problems, such as a three-level atom \cite{hoffmannBenchmarkingSemiclassicalPerturbative2019} where the decoupling scheme can be extended so that up-hill and down-hill energy transfer is always well accounted for without drawing energy from vacuum fluctuations. 

It would also be of interest to combine the DC-MF formalism with existing implementations of classical field algorithms such as FDTD and FEM in order to render such methods compatible with quantum models of matter, where it should be noted that the optical mode-resolved equations of motion employed in our work can straightforwardly be transformed to the more commonly used position basis. Lastly, it is noteworthy that encouraging results have been obtained through quasiclassical mapping Hamiltonian methods for the same cavity-embedded atom problem such as considered here \cite{SallerBenchmarkingQuasiclassicalMapping2021}. Given that MF dynamics can be formulated as a member of such methods, our work could provide helpful guidance for further improvements.

\section*{Acknowledgement}
This work was supported by the National Science Foundation's MRSEC program (DMR-1720319) at the Materials Research Center of Northwestern University.
 
\bibliography{MixQC_bib}

\end{document}